\documentclass[twoside]{ilcws07}
\usepackage[latin1]{inputenc}
\usepackage{graphicx}
\usepackage{epsfig}
\usepackage{wrapfig,rotating}
\usepackage{amssymb,amsmath,array}
\usepackage{epstopdf}
\begin{document}
\title{
SiLC simulation status report} 
\author{M.A. Vos  \\
\vspace{.3cm}\\
IFIC (U. Valencia/CSIC) \\
Apartado de Correos 22085, \\
E-46071 Valencia, Spain \\
}

\maketitle

\begin{abstract}
The SiLC - Silicon for the Linear Collider - collaboration aims to develop silicon detector technology for tracking in the international linear collider experiments. The R \& D programme involves a substantial effort in simulation of the response of detector designs. In this contribution, an overview of ongoing efforts in this area is given.
\end{abstract}

\section{Introduction}

The SiLC R \& D collaboration~\cite{silc} for the International Linear Collider involves over 200 people from 22 institutes. SiLC members are from three continents, and pertain to the three detector concepts: LDC, SiD and GLD. The aim of the collaboration is to develop silicon detector technology for tracking in the ILC experiments. Core activities include Front End chip design for micro-strip detectors, sensor development, (hardware) alignment and mechanics.

Monte Carlo simulation is an indispensable tool to guide the detector design.  In the past, SiLC has played a mayor role in the development and maintenance of valuable fast simulation packages. Well-known examples are the Simulation a Grande Vitesse (SGV)~\cite{sgv}. More recently, the LicToy package - described elswhere in these proceedings~\cite{lictoy} - has been developed. Today, the attention of the SiLC simulation team is shifting towards {\em full simulation}: detailed Monte Carlo studies of the detector response. 

Due to lack of space, several important areas of work will not be discussed in this contribution. Examples are the active line of development centered at Charles University in Prague towards a versatile digitization package for micro-strip detectors, and the investigation of the reconstruction of non-prompt tracks in the SiD concept.

In this contribution, an overview is given of recent progress in the SiLC simulation studies. In separate sections, three specific items are highlighted. The first two deal with two aspects of tracking in the very forward region of the ILC experiment - a complex area that has received relatively little attention. In section~\ref{sec-forward-bkg} estimates are presented of the hit density in the Forward Tracking Disks due to machine background. Section~\ref{sec-forward-mat} deals with the impact of the material budget on the momentum resolution in the same region. The third aspect of tracking discussed in this contribution is the estimate of extrapolation uncertainties in the innermost part - the Silicon Intermediate Tracker - of the central tracker.

\section{Background levels in the forward region}
\label{sec-forward-bkg}

In an $e^+ e^- $ collider like the ILC there is one major source of machine-induced background: incoherent pair creation off beamstrahlung photons. An accurate generator for this complex process is available~\cite{guineapig} and full detector simulations have been performed, primarily for the vertex detector. The effect on other sub-detectors - in particular the TPC~\cite{vogel} - is actively being investigated. In this section, the background levels in the forward tracker are discussed briefly.  

The inner radius of the forward tracking disks and therefore the angular coverage of the tracker is limited by a region - the accumulation zone - of extremely high background intensity. The radial extent of this zone at different distances along the beam axis is studied using an approach based on the work of reference~\cite{rimbault}. The trajectory of the electrons and positrons is represented by a helix in a solenoidal field. Interactions with material are neglected.

For nominal machine parameters and a 4 Tesla magnetic field the accumulation zone is found to extend to a radius of less than 15 mm for z-positions below 320 mm. For alternative parameter sets of the final focus system, the accumulation zone reaches much larger radii. For the low-power (high-luminosity) option the radius at the first (at z= $ \pm $ 20 cm) and second (z= $ \pm $ 32 cm) disk of the LDC design become, respectively, 16 (21) mm and 24 (30) mm. Thus, the envisaged inner radius of the active region of the innermost LDC forward tracker disks (37.5 mm) leaves only approximately 7 mm for the beam pipe and the detector services.

\begin{wrapfigure}{r}{0.5\columnwidth}
\centerline{\includegraphics[width=0.45\columnwidth]{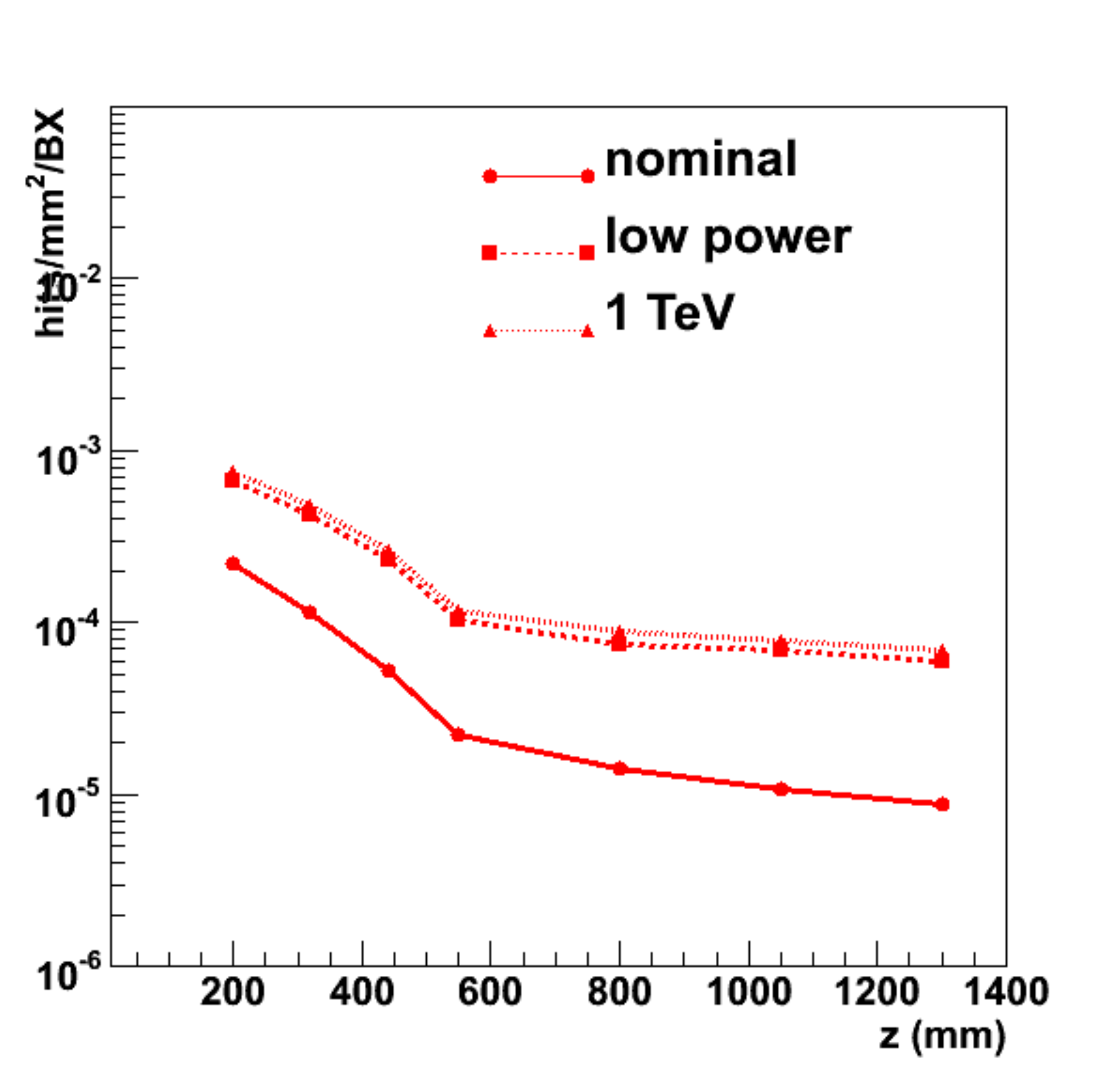}}
\caption{The hit density due to machine background in the LDC forward tracker.}\label{fig-forward-bkg}
\end{wrapfigure}

A quantitative estimate of the pair background hit density in the forward tracker of the LDC geometry is obtained using a full GEANT4 simulation of the response of the detector to the pairs generated with GuineaPig~\cite{guineapig}. The results for the seven disks of the LDC geometry are shown in figure~\ref{fig-forward-bkg}. The three solid curves indicate the hit density in units of hits $/mm^{2}/$ BX for a 500 GeV collider with nominal final focus parameters. On the innermost two disks hit density reaches $2 \times 10^{-4}$, comparable to the level in the outermost layer of the vertex detector. For the outermost disks the background level is an order of magnitude lower. 

The dashed curve corresponds to a variation of the final focus parameters known as the low power option. The dotted curve respresents a 1 TeV linear collider with nominal parameters. The background level is quite sensitive to both parameters: both result in a factor two increase of the background level with respect to the nominal parameter set throughout the length of the forward tracker.

\section{Impact of the material budget in the forward region}
\label{sec-forward-mat}

The ILC detector concepts aim for a superb transverse momentum resolution. Compared to the central tracker, the momentum resolution in the very forward region is inevitably degraded by the less favourable orientation of the magnetic field. Only through the use of state-of-the-art instrumentation can the degradation of the performance towards small polar angle be limited.

\begin{wrapfigure}{r}{0.5\columnwidth}
\centerline{\includegraphics[width=0.45\columnwidth]{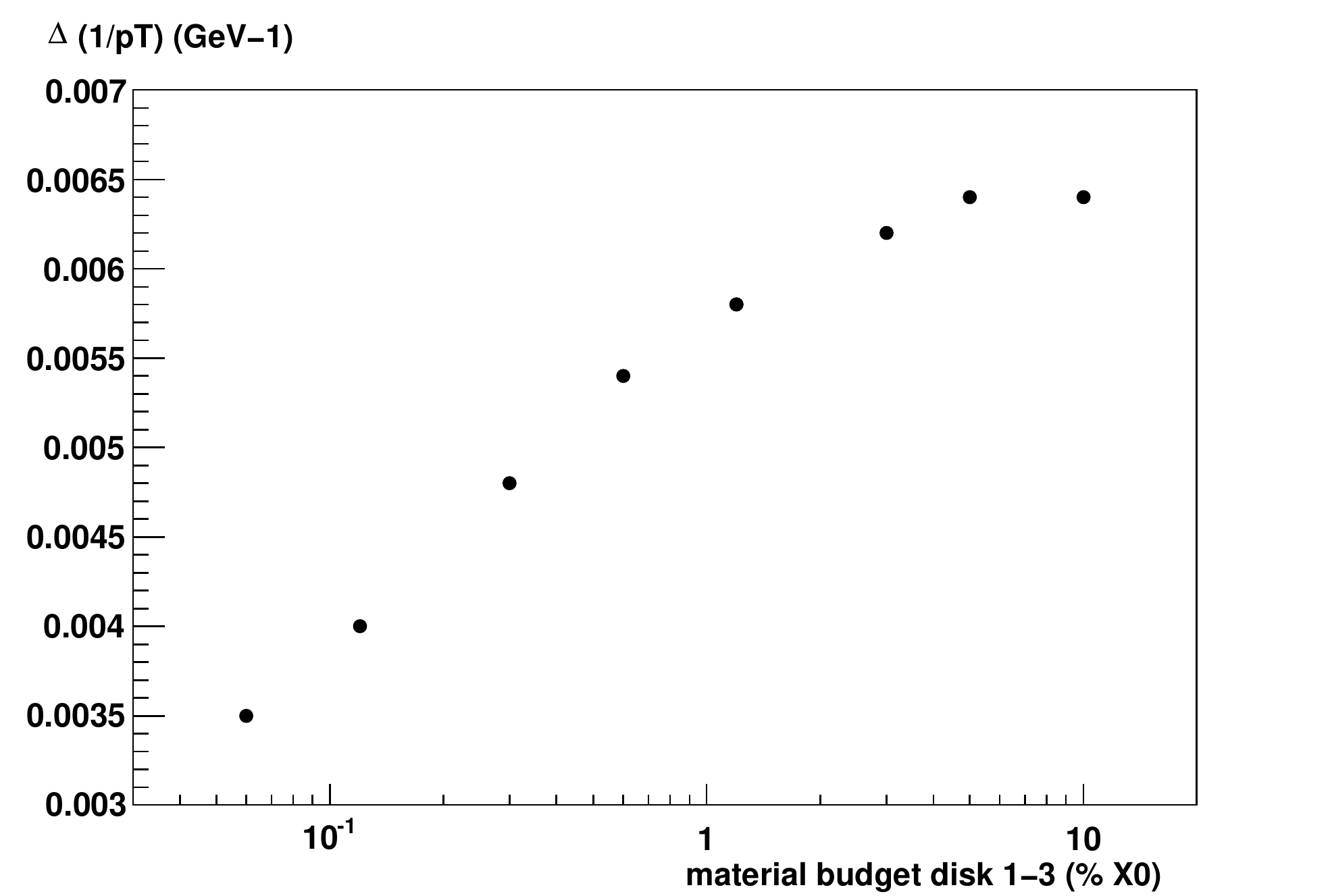}}
\caption{The transverse momentum resolution for 1 GeV tracks in the LDC forward tracker versus material budget in the innermost disks.}\label{fig-forward-res}
\end{wrapfigure}

In this section the performance of two possible detectors is compared. Both options correspond to the nominal LDC layout and identical assumptions are made regarding the space point resolution: 5 $ \mu m R \phi $-resolution for the  three innermost disks, 10 $ \mu m $ for the four outermost disks. The two setups differ the assumption made for the material budget. The first set-up corresponds to a rather conservative estimate of 1.2 \% $ X_{0} $/disk for the first three disks and 0.8 \% $X_{0}$ for the four outermost disks. For the second set-up the material in the three disks is reduced by a factor 10.

An estimate of the momentum resolution is obtained using a Kalman Filter track fit on a geometry representing the LDC layout for a track at a polar angle of $20^o$. As a cross-check, two independent implementations of the fitting algorithm were used, LicToy~\cite{lictoy} and the CMS track fit. Both packages are found to yield compatible results.

The results for a range of particle momenta are parametrized as the quadratic sum of a constant (resolution) term and a (multiple scattering) term proportional to $ 1/p_{T}$:

 $\sigma (p_T)/p^2_T               = 2.0 \times 10^{-4} \oplus 5.8 \times 10^{-3}/p_T $  		
, if FTD 1-3 have 1.2 \% $X_{0}$/disk

$ \sigma (p_T)/p^2_T   = 1.8 \times 10^{-4} \oplus 4.0 \times 10^{-3}/p_T $
, if FTD 1-3 have 0.12 \% $X_0$/disk

Clearly, the reduction of the material budget in the challenging set-up leads to an improved momentum resolution performance. As expected, the effect is particularly signficant at low momentum, where multiple scattering has the largest impact. In figure~\ref{fig-forward-res} the transverse momentum resolution for charged particles with a transverse momentum of 1 GeV and a polar angle of $20^o$ is shown for a series of assumptions on the material budget of the three innermost forward tracking disks.

The improvement of the momentum resolution may seem small in comparison to the effort of reducing the material. It should be noted, however, that there are very few degrees of freedom in the FTD design. While the resolution for high-momentum tracks can be improved signficantly by a better space point resolution, the material is the principal handle on the resolution of low-momentum tracks. 

\section{Pattern recognition in the central tracker}
\label{sec-sit-patrec}

To meet the ILC requirement of efficient and clean reconstruction of charged particle tracks, the various sub-detectors should have excellent pattern recognition capabilities. Currently, the SiLC simulation team is investigating the constraints on the tracker design that derive from this requirement. 

One key ingredient to pattern recognition is the precision with which track candidates can be extrapolated to the { \em next } layer. The uncertainty on the extrapolated position determines - together with the hit error - how many unrelated hits will be compatible with the track candidate. The extrapolation precision depends on the precision of the track parameter estimate and on the distance over which the track is extrapolated. For low momentum tracks, the amount of material in the last measurement layer is furthermore important.

\begin{wraptable}{r}{0.5\columnwidth}
\centerline{\begin{tabular}{|l|r|}
\hline
VXD parameters  & R $ \phi $ uncertainty ( $ \mu m $ ) \\ \hline
LDC nominal            &  61 $ \oplus $ 114  / $ p_T $ \\
material $ \times $ 2  &  61 $ \oplus $ 166  / $ p_T $ \\
resolution $ \times $ 2  &  122 $ \oplus $ 117 / $ p_T $  \\
4-layer VXD  &  105 $ \oplus $ 134   / $ p_T $\\
All-together  &  211 $ \oplus $ 199 / $ p_T $  \\  \hline
SIT radius  & R $ \phi $ uncertainty ( $ \mu m $ ) \\ \hline
R = 12 cm & 27 $ \oplus $ 47 / $ p_T $ \\
R = 20 cm & 108 $ \oplus $ 215 / $ p_T $ \\
R = 30 cm & 279 $ \oplus $ 624 / $ p_T $ \\
\hline
\end{tabular}}
\caption{The $ R \phi $ precision of the prediction position on the first Silicon Intermediate Tracker layer under a variety of assumptions for the detector layout.}
\label{tab:pattern}
\end{wraptable}

In the following, a track is ``grown'' by an iterative process of extrapolation to the next layer, a search for compatible hits, and update of the trajectory parameters. The track search is ``inside-out'', i.e. seeds are created in the VXD. In the first detector layers, the track candidate is only weakly constrained. As a result of the excellent space point resolution and tight material budget, the prediction on the next vertex detector layer is quite precise. Assuming a 2 $ \mu m $ two-dimensional space point resolution and a material of 0.12 \% $ X_0 $  the uncertainty in the predicted position when extrapolating a track candidate consisting of three VXD measurements to the outermost VXD layers is of the order of 5 $ \mu m $.

For the next step - the extrapolation of a 5-point track to the intermediate tracker layers - the extrapolation distance is signficantly larger. In the various detector concepts the Intermediate Tracker has an innermost layer at a radius 9 cm (GLD), 16 cm (LDC) or 22 cm (SiD). In the LDC layout the extrapolation precision at the innermost SIT layer is given by: $ \sigma ( R \phi ) = 61 \oplus 114 / p_T $, whereas $ \sigma ( z) = 7 \oplus 43 /p_T $. The larger $ R \phi $-uncertainty reflects the weakly constrained transverse momentum of the track. In table~\ref{tab:pattern} the $ R \phi $ extrapolation precision is listed for a variety of assumptions on the VXD performance and VXD-SIT distance. 

\section{Conclusions}
The SiLC collaboration is actively pursuing a number of simulation studies. In this contribution, three highlights have been presented. In the coming year these studies should allow to determine the requirements of the intermediate central tracker and the forward tracker.

\begin{footnotesize}

\end{footnotesize}
\end{document}